\documentstyle[11pt,newpasp,psfig]{article}

\begin{document}

\title{Solar Surface Magneto-Convection}

\author{Robert F. Stein}
\affil{Physics and Astronomy Department, Michigan State University,
    East Lansing, MI 48824}

\author{{\AA}ke Nordlund}
\affil{Niels Bohr Institute for Astronomy, Physics, and Geophysics
University of Copenhagen, DK}

\setcounter{footnote}{0}

\begin{abstract}

Magneto-convection simulations on meso-granule and granule scales near the 
solar surface are used to study small scale dynamo activity, the emergence 
and disappearance of magnetic flux tubes, and the formation and evolution 
of micropores.  

From weak seed fields, convective motions produce highly intermittent 
magnetic fields in the intergranular lanes which collect over the boundaries 
of the underlying meso-granular scale cells.  Instances of both emerging 
magnetic flux loops and magnetic flux disappearing from the surface occur 
in the simulations.  We show an example of a flux tube collapsing to kG 
field strength and discuss how the nature of flux disappearance can be 
investigated.  Observed stokes profiles of small magnetic structures are 
severely distorted by telescope diffraction and seeing.

Because of the strong stratification, there is little recycling of plasma 
and field in the surface layers.  Recycling instead occurs by exchange 
with the deep layers of the convection zone.  Plasma and field from 
the surface descend through the convection zone and rise again toward the 
surface.  Because only a tiny fraction of plasma rising up from deep in
the convection zone reaches the surface due to mass conservation, little 
of the magnetic energy resides in the near surface layers.  Thus the dynamo 
acting on weak incoherent fields is global, rather than a local surface dynamo.  

\end{abstract}




\section{Introduction}

We model magneto-convection in a small domain near the solar surface
by solving the partial differential equations for mass, momentum and
internal energy conservation and the induction equation for the
vector potential.  Our goal is to make a realistic representation of
the solar surface (Stein \& Nordlund 2000).  Our domain is 6
$\times$ 6 Mm horizontally and extends from the temperature minimum,
0.5 Mm above continuum optical depth unity, to 2.5 Mm below the
visible surface, using a grid 253 $\times$ 253 $\times$ 163 grid
points, which gives a horizontal resolution of 25 km and a vertical
resolution of 15 km near the surface increasing to 35 km at the
bottom.  The initial state was a snapshot of non-magnetic solar
convection on which was imposed a uniform magnetic field: either a
horizontal seed field of 1 G or 30 G, or a vertical field of 400 G.
The boundary conditions are periodicity in horizontal directions,
open boundaries for the fluid in the vertical direction, and the magnetic
field at the top tends toward a potential field.  The magnetic bottom
boundary condition for horizontal seed fields was that inflows advect
in horizontal 1G or 30G field, while in outflows the vector potential
is advanced in time from the induction equation with the current
calculated using spline derivatives with the cubic spline condition
that the third derivative is continuous.  The magnetic bottom
boundary condition for the vertical field was that the field tend
toward the vertical.  Rotation and coriolis forces are neglected,
because on this small scale of mesogranulation, with a depth of only
3 Mm, the flows don't feel the rotation.

\section{Three-dimensional Effects on the Mean Structure} 
\label{sec:3Dvs1D}

The mean atmospheric structure is different in 3D than in 1D.  There
are two major reasons for this.  First is a 3D radiative transfer
effect.  The temperature is very inhomogeneous near the surface of
stars.  In cool stars the dominant opacity source is H$^{-}$, which
is very temperature sensitive -- hot gas is much more opaque than
cool gas.  As a result, we only see the cool gas.  Therefore, the
average temperature, for a given effective temperature, is higher in
3D models than in 1D models and thus the scale height is larger and
the atmosphere more extended.  Second, convective motions produce a
turbulent pressure which also contributes to the support of the
atmosphere.  The net effect of these two phenomena together is that
atmosphere is raised about one full scale height in the 3D models
compared to the 1D models.

\begin{figure*}[!htb]
\centerline{\psfig{file=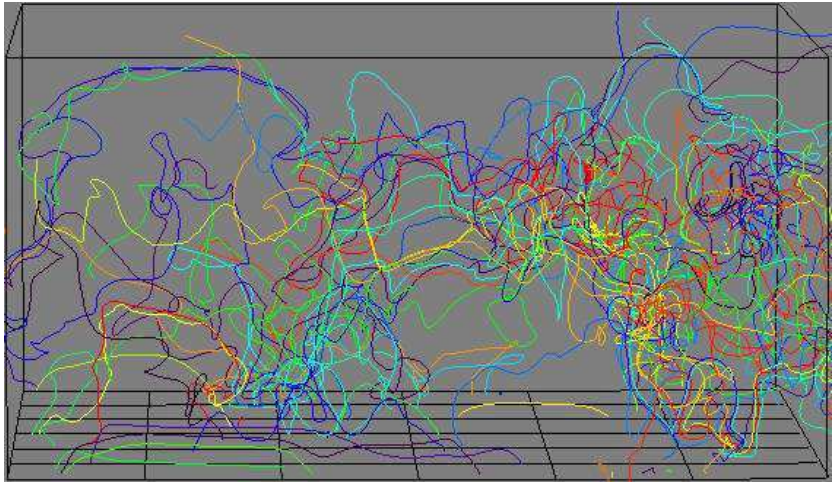,width=11cm}}
\caption{Selected magnetic field lines in a snapshot viewed from the 
side.  Horizontal field is advected in by ascending flow at the bottom 
boundary.  It gets advected, stretched and twisted by the convective 
motions.
}
\label{fieldlines30Gside}
\end{figure*}

\section{Magneto-Convection} \label{sec:magconv}

\subsection{Magnetic Field Organization} \label{sec:organization}

\begin{figure*}[tb]
\centerline{\psfig{file=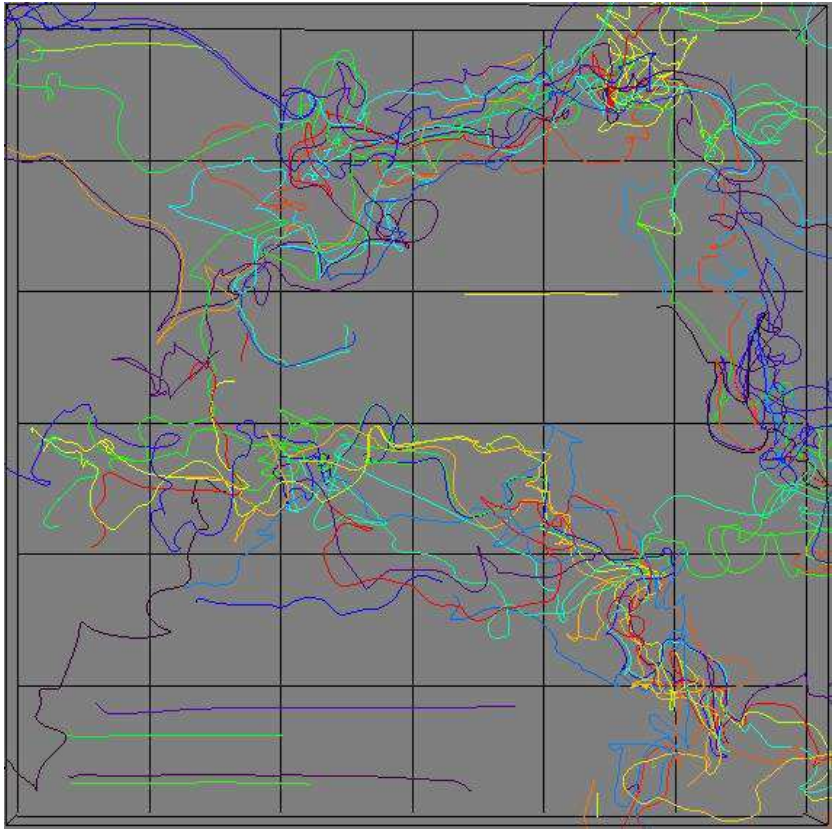,width=11cm}}
\caption{The same magnetic field lines as in Fig 
1, viewed from the top.  Squares at the bottom are 1 $\times$ 1 Mm.
The magnetic field is advected into the computational domain by upflows
in the interiors of the mesogranules and is swept out of the granules 
and mesogranules into the downflows in the mesogranule boundaries.
}
\label{fieldlines30Gtop}
\end{figure*}

\begin{figure*}[!htb]
\centerline{
\psfig{file=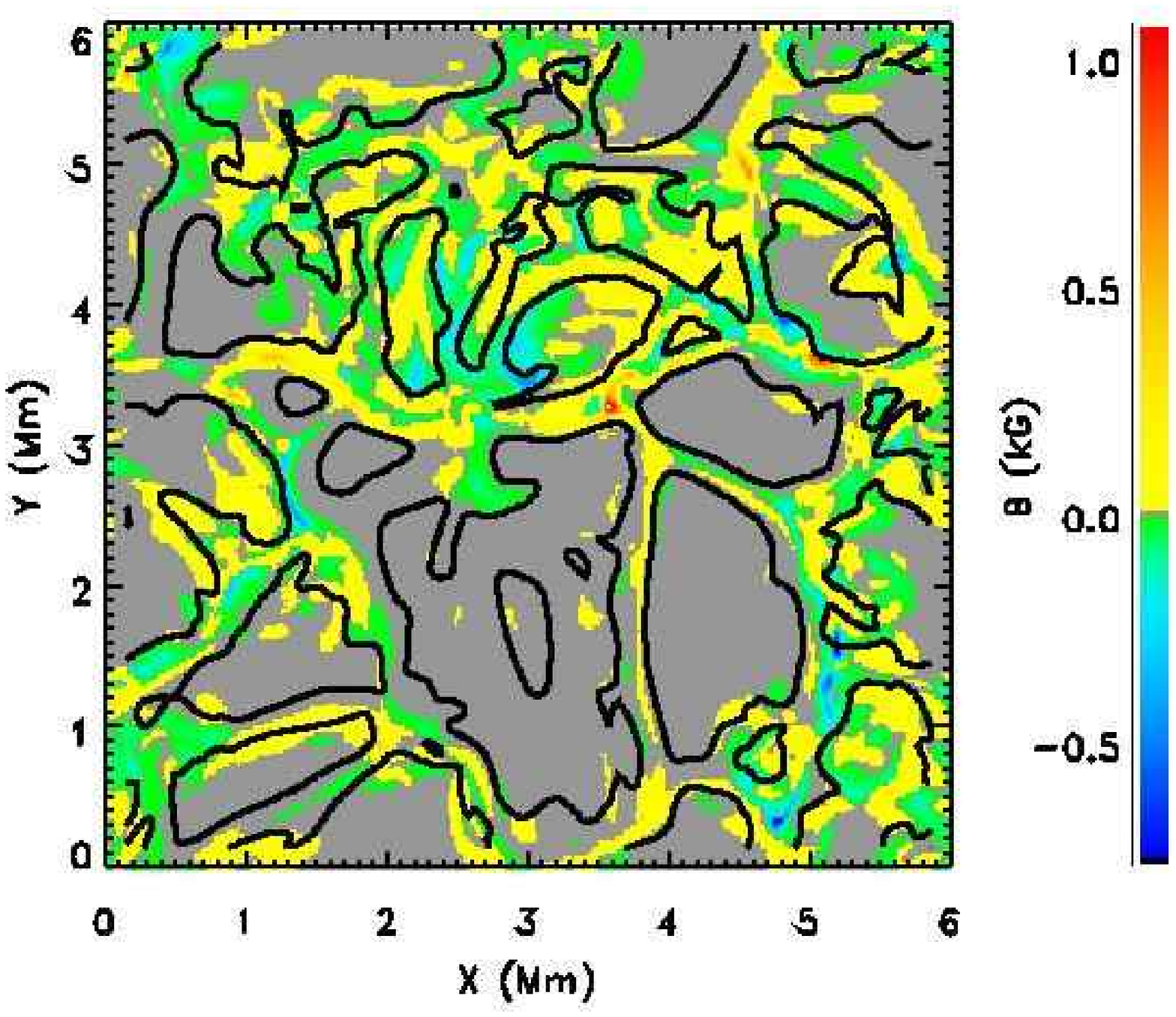,width=4.4cm}
\psfig{file=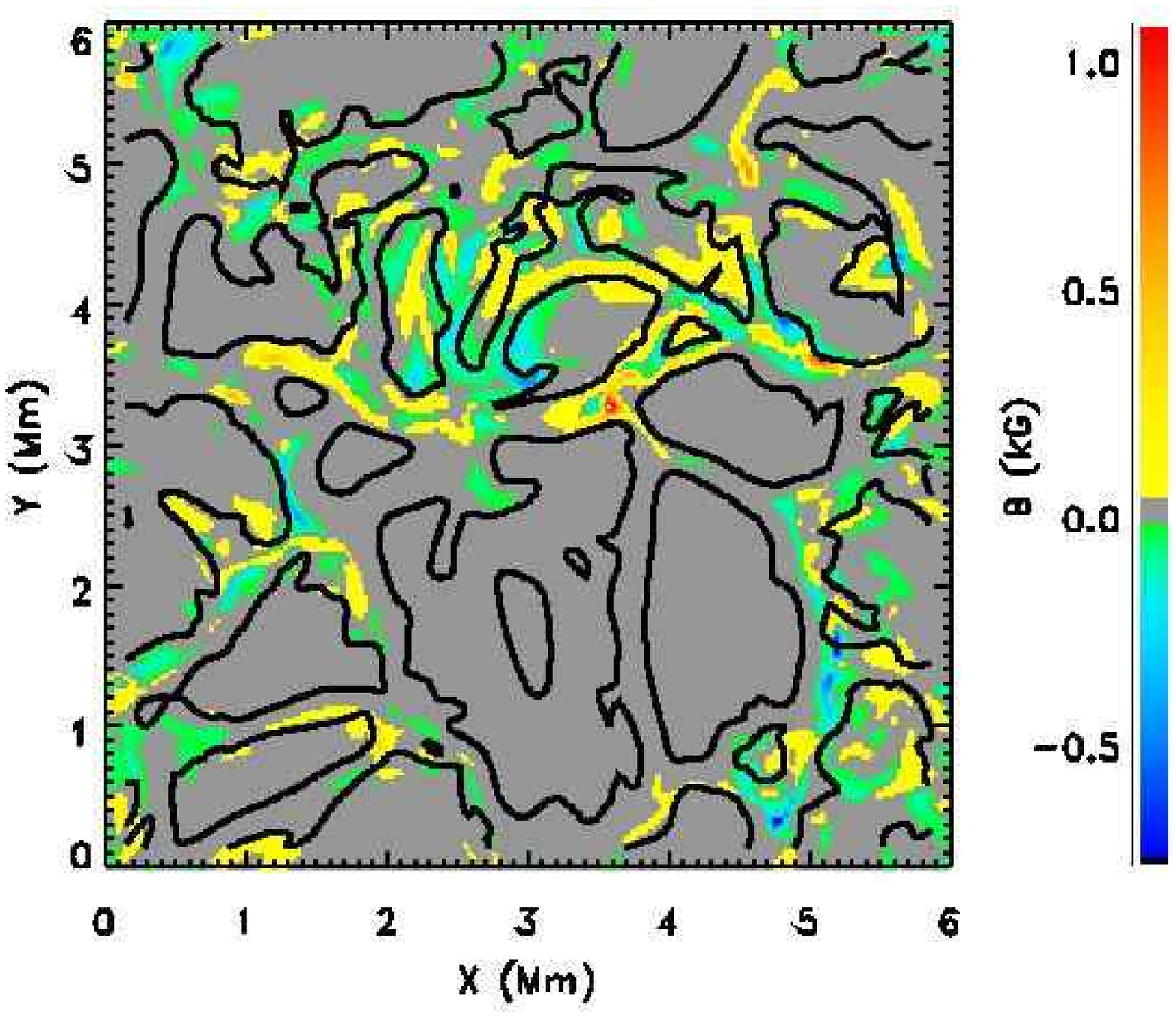,width=4.4cm}
\psfig{file=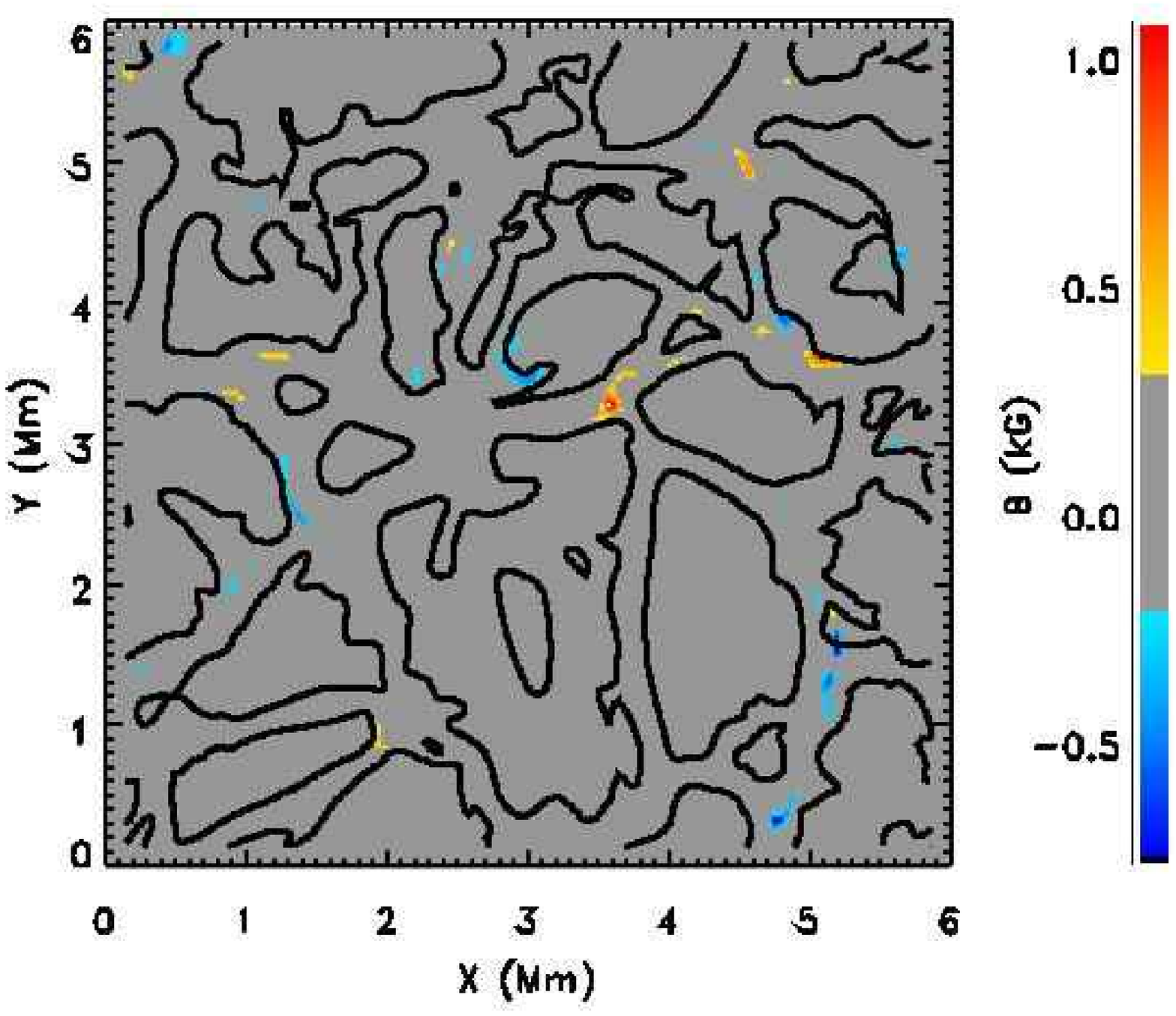,width=4.4cm}
}
\caption{Image of magnetic field with superimposed zero velocity 
contours to outline the granules for the case of a 30 G uniform 
horizontal seed field.  Field magnitudes less than 3, 30 and 300 G 
respectively are shown in gray.  The magnetic field is concentrated 
into the intergranular lanes.  It is highly intermittent, with strong 
fields occupying a tiny fraction of the total area.
}
\label{Bz_3_30_300}
\end{figure*}

For the case of horizontal field advected in from the bottom, to date
we have run an almost two hour sequence with a 1G field and a 30
min.  sequence with a 30 G field.  Figure \ref{fieldlines30Gside}
shows selected magnetic fieldlines for one snapshot for the 30G
case.  At the bottom there is horizontal magnetic field being
advected into the computational domain by ascending fluid.  Higher up
are the twisted filed lines produced by the turbulent convective
motions.  It looks pretty chaotic.  However, viewed from the top one
sees that the the field is actually organized on a larger 
scale than the granulation (Fig.~\ref{fieldlines30Gtop}).  It is
being organized on the scale of the underlying mesogranules.  The
horizontal field is entering in the interiors of the mesogranules
where there are the upflows.  The overlying magnetic field has been
twisted and stretched and swept out of the granules and mesogranules
into the boundaries of the mesogranules, where the plasma is descending.  
This produces a highly
intermittent field with a stretched exponential distribution of field
strengths, as has been found in the Boussinesq calculations of
Cattaneo (1999) and as observed by, e.g. Harvey \& White (1999)
and Hagenaar (2001).  This means that the stronger the field the
tinier the fraction of the area it occupies.  Fields stronger than 3
G, fill all the intergranular lanes and exist even inside some of the
granules.  Fields that are stronger than 30 G have been swept out of
the granules into the intergranular lanes and even some the
intergranular lanes have no field stronger than 30 G.  Finally,
fields stronger than 300G occur hardly anywhere in the domain
(Fig.~\ref{Bz_3_30_300}).

\subsection{Flux Emergence and Disappearance}

\begin{figure}[!htb]
\centerline{\psfig{file=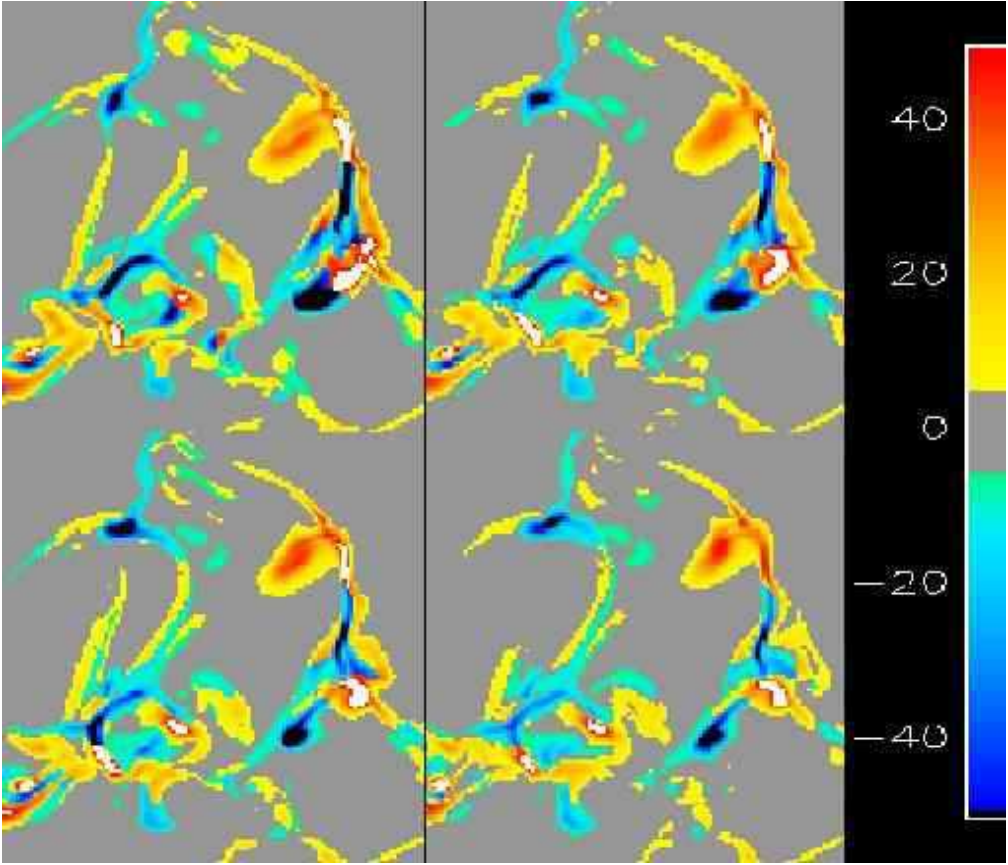,width=\linewidth}}
\caption{Four snapshots of the surface magnetic field magnitude at 
10 sec. intervals (TL, TR, BL, BR) showing the emergence and 
footpoint  separation of a magnetic loop on the right hand side 
and the disappearance of magnetic flux on the left hand side.
}
\label{flux_emer_disp}
\end{figure}

In these simulation we see examples of flux emergence, merging,
fragmentation, and cancellation.  Figure~\ref{flux_emer_disp} is a
sequence of images at 10 sec. intervals showing a flux loop emerging
on the right hand side and its foot points separating.  New flux
emerges sometimes inside granules, sometimes at their edges, and
sometimes in intergranular lanes.  Flux emerging inside granules is
quickly swept into intergranular lanes by the diverging upflows of
the granulation.  Inside the intergranular lanes, the magnetic field
is advected by horizontal flows (cf. Stein \& Nordlund 1998).
At the same time, on the left side, oppositely (vertical component)
directed flux is coming together and partially annihilating.  We
still need to analyze this process to see what is actually
occurring.

\subsection{Stokes Profile Observations}

\begin{figure*}[!htb]
\centerline{
\hspace*{0.3cm}
\psfig{file=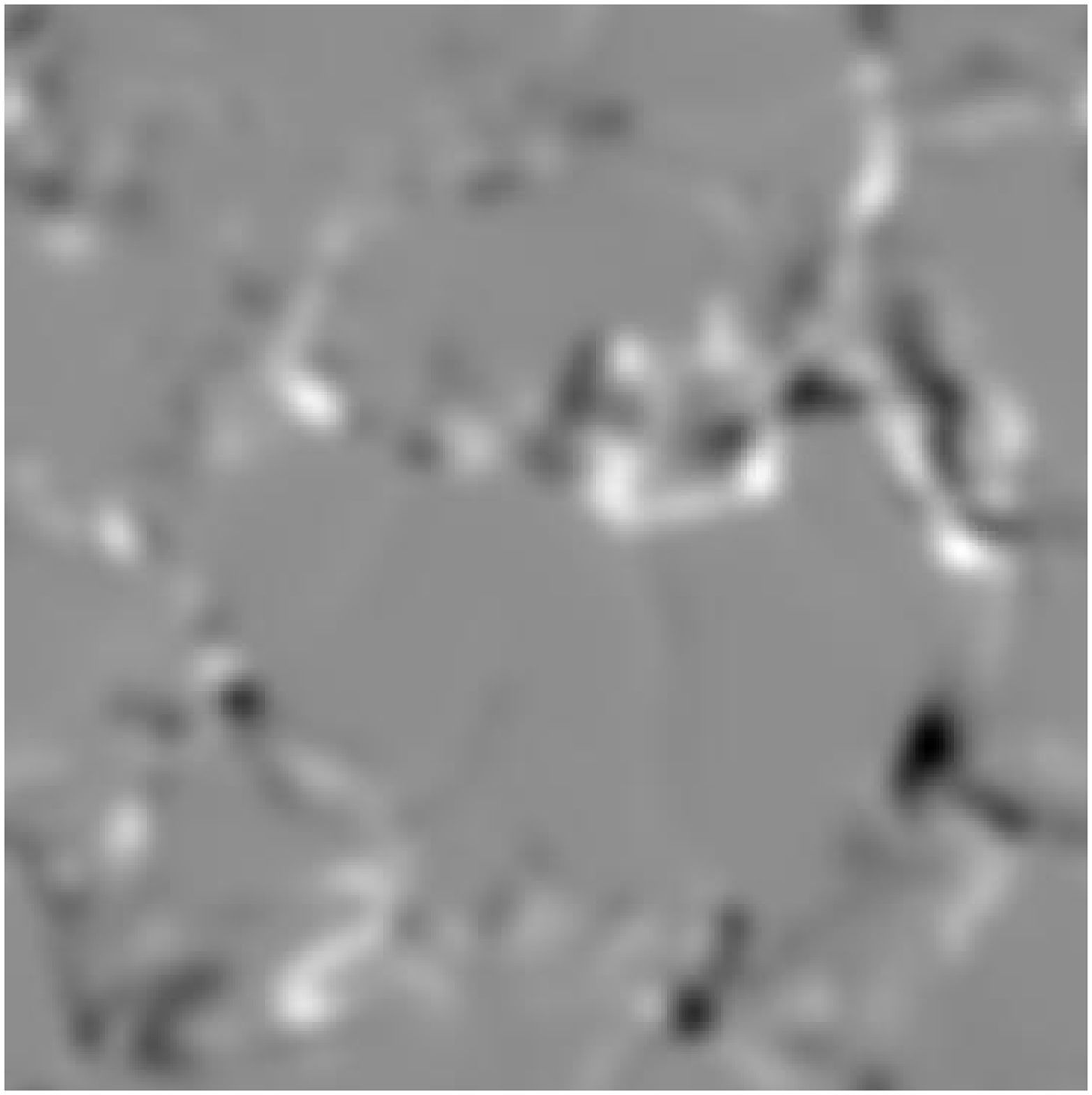,width=4.8cm}
\psfig{file=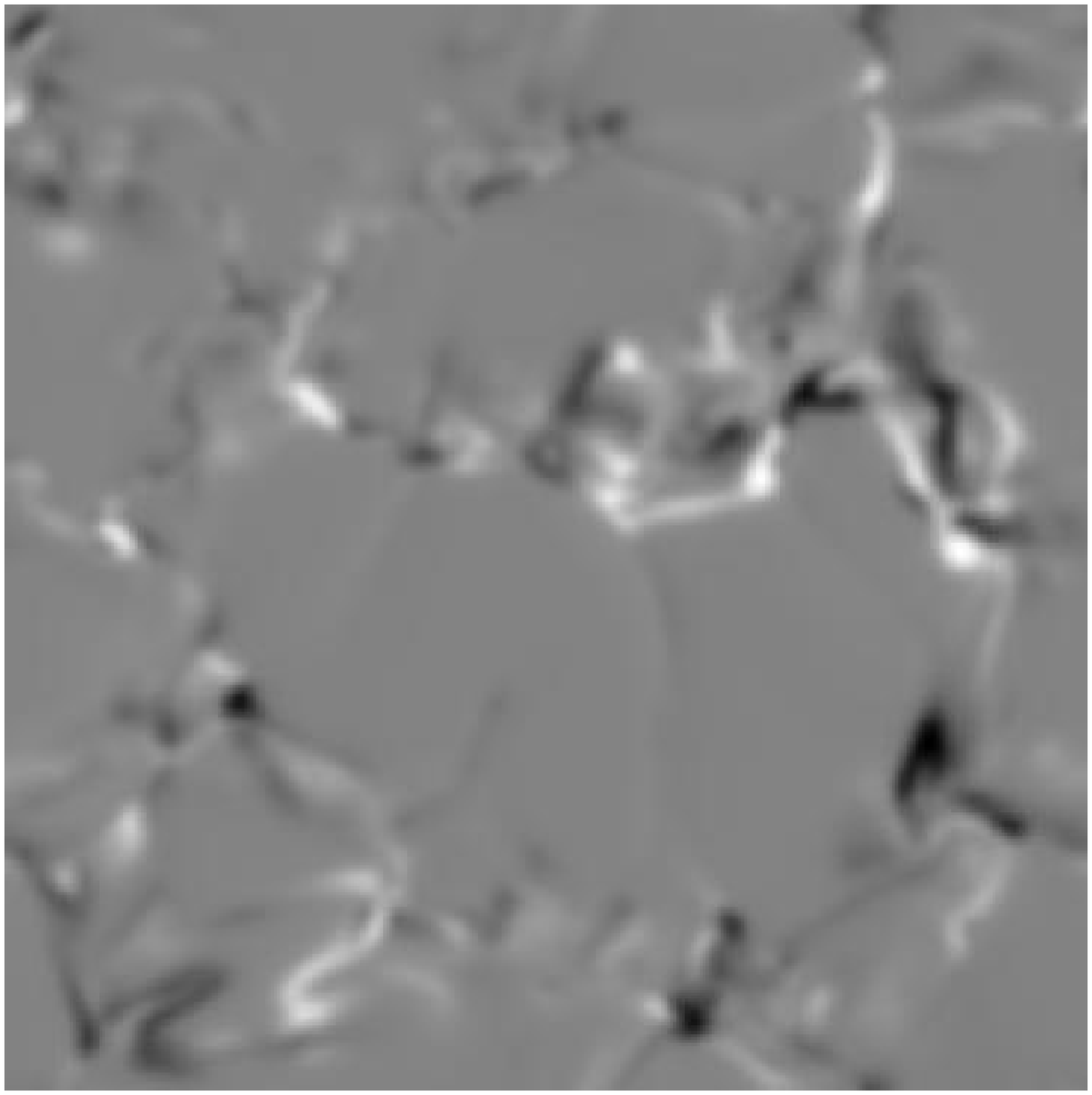,width=4.8cm}
\psfig{file=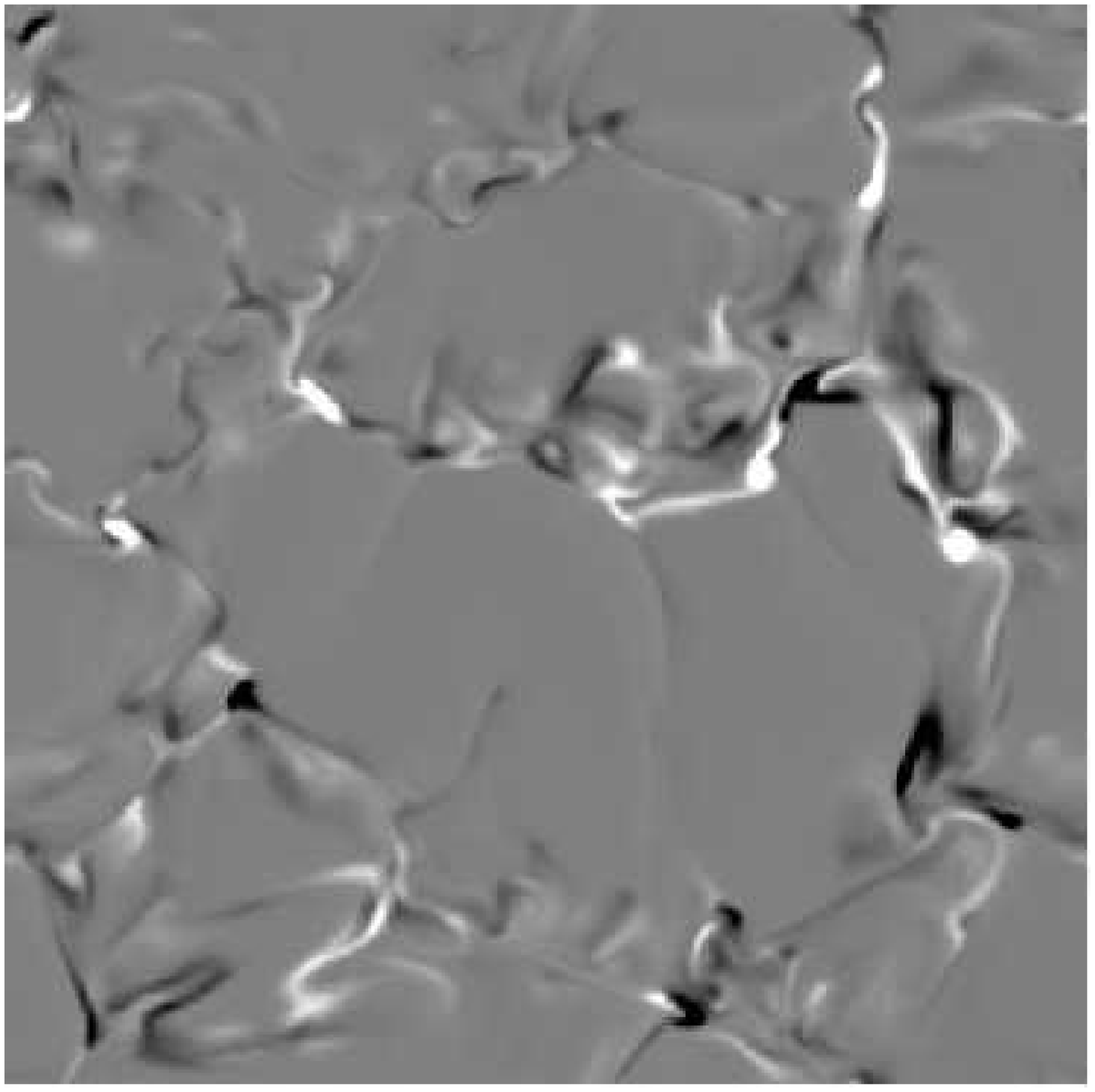,width=4.8cm}
}
\caption{Stokes V image as it would be observed with perfect seeing at
the old 47 cm Swedish Solar Telescope on La Palma (left), the new 97 cm
telescope (center), and the image as obtained from the simulation (right).
}
\label{stokes_images}
\end{figure*}

\begin{figure}[!htb]
\centerline{
\psfig{file=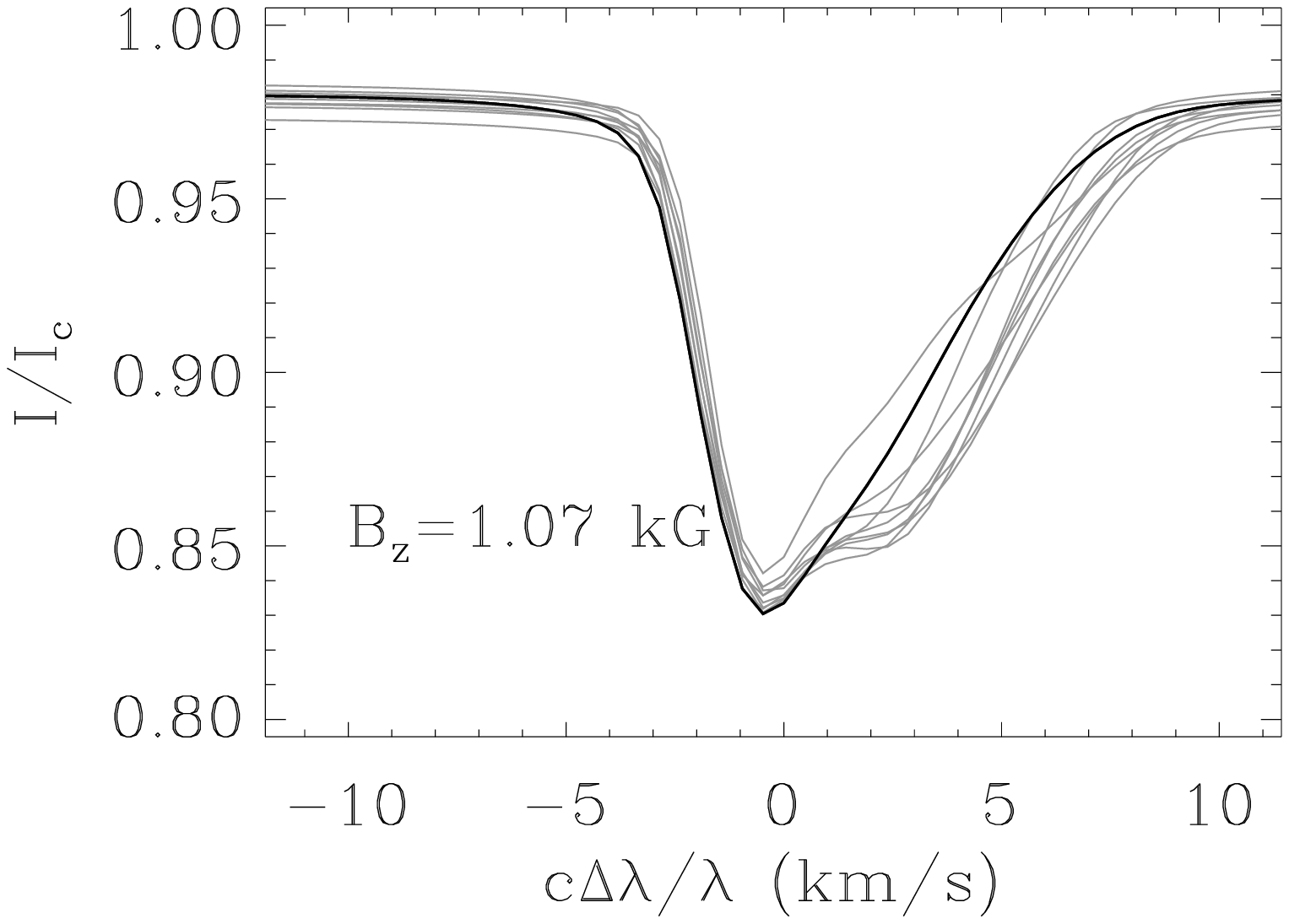,width=5.7cm}
\psfig{file=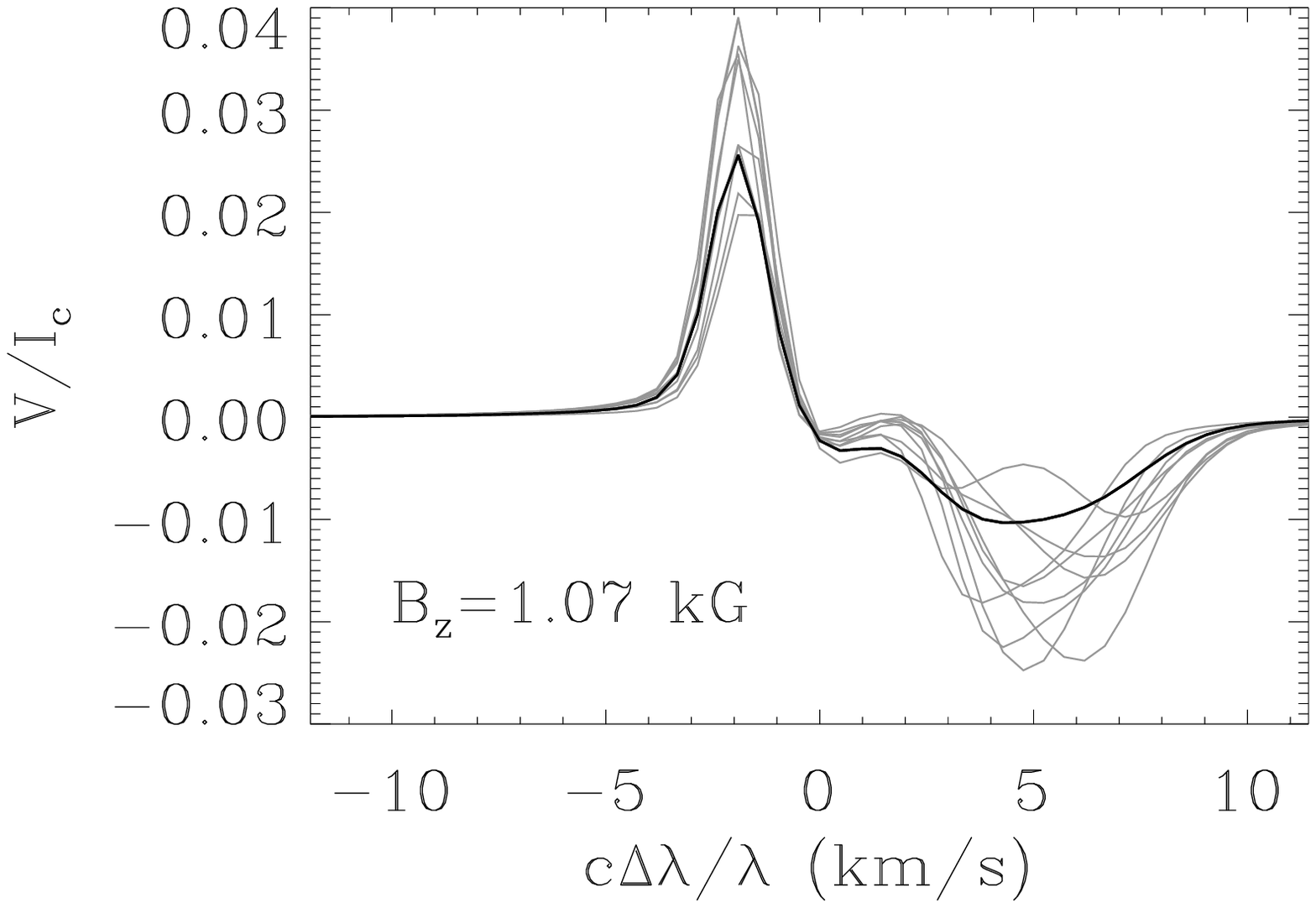,width=5.7cm}
}
\centerline{
\psfig{file=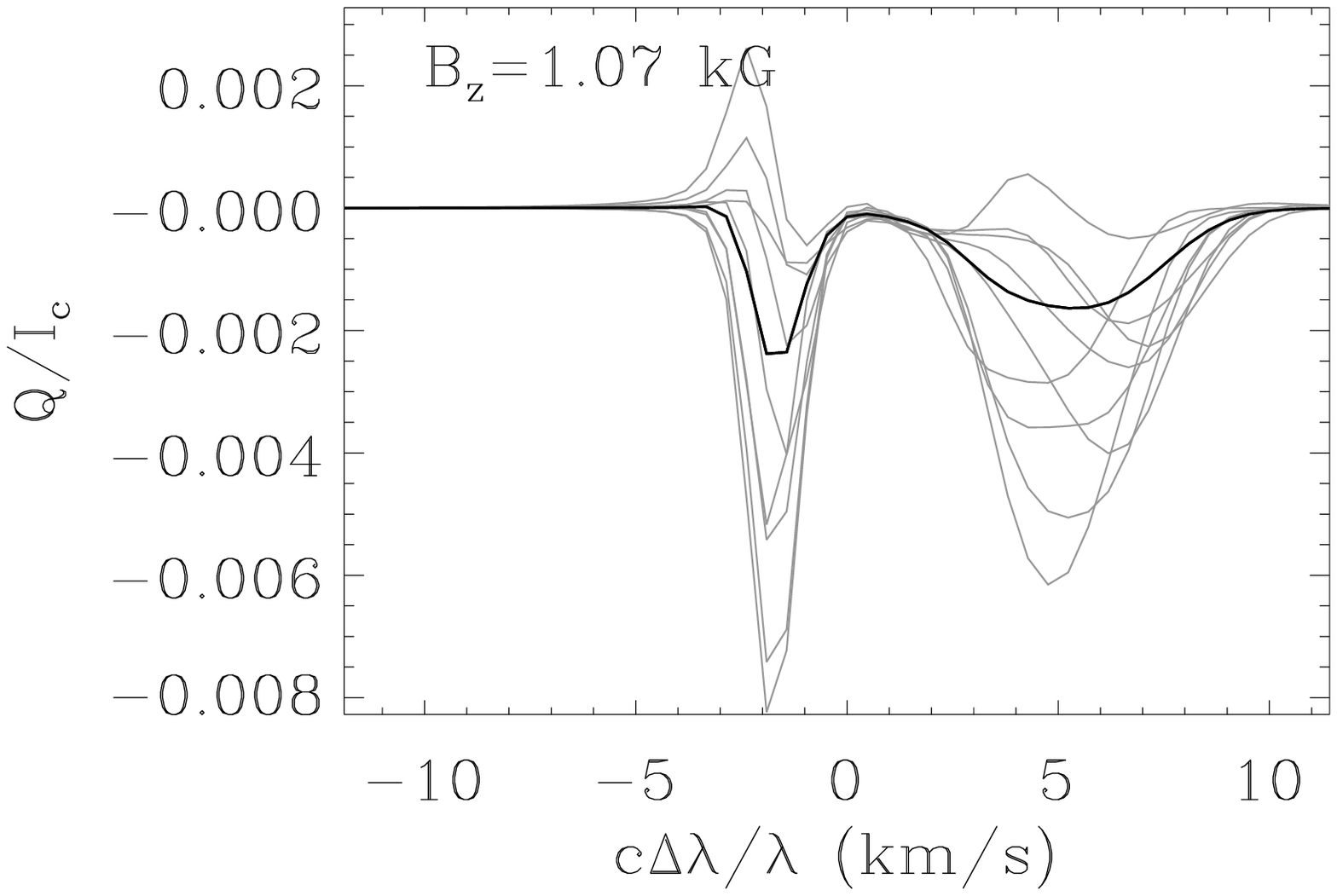,width=5.7cm}
\psfig{file=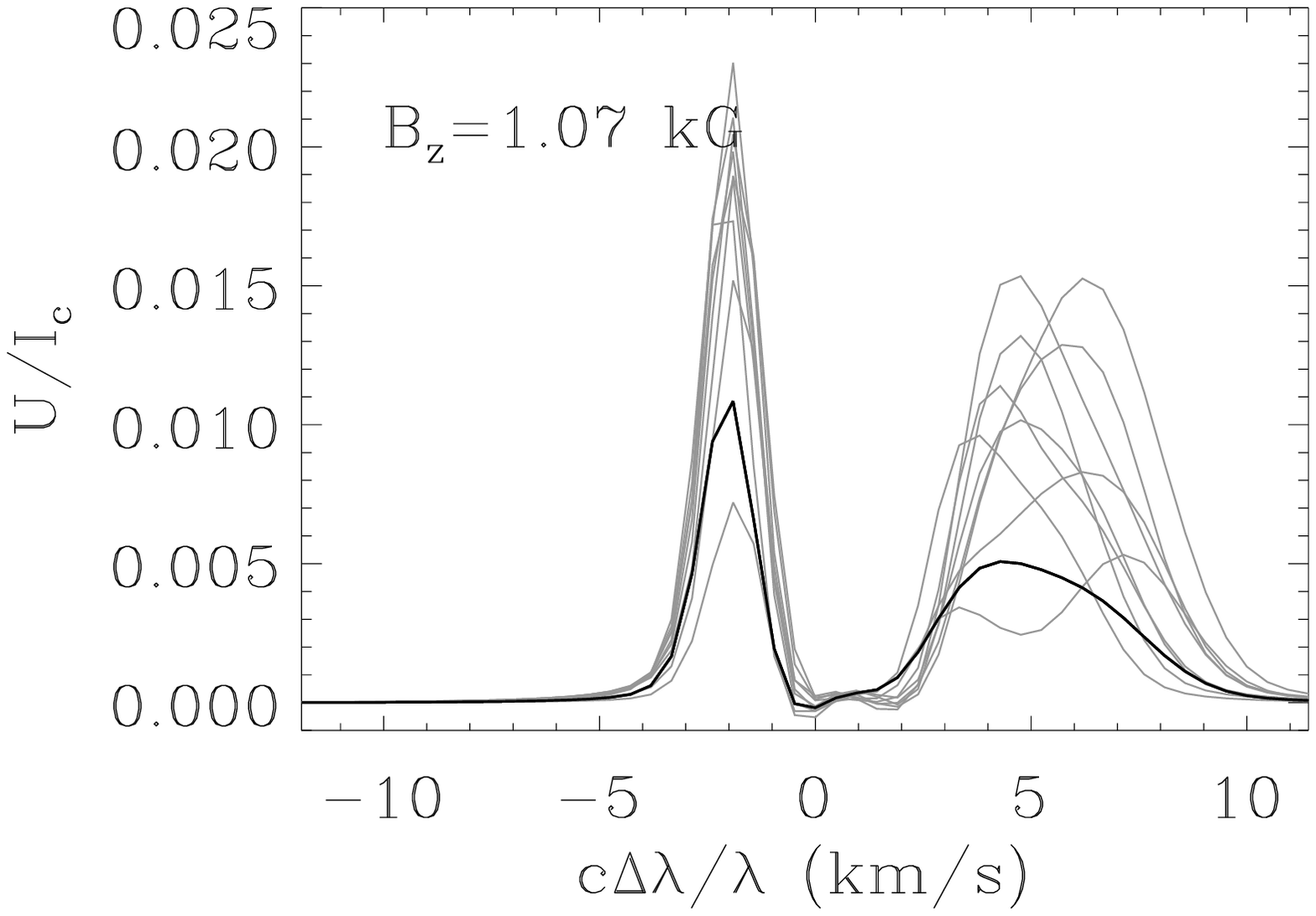,width=5.7cm}
}
\caption{Stokes I,V,Q,U profiles for FeI 6302 as would be observed by
the 97 cm Swedish Solar Telescope on La Palma
with perfect seeing (thick lines) compared with the profiles for 
9 individual grid cells covering 75 $\times$ 75 km region around the 
central point (gray lines).  The profile is degraded by the finite 
telescope resolution with respect to what actually occurs in the 
simulation.  
}
\label{stokes_profiles}
\end{figure}

Resolution has a tremendous impact on how one interprets
observations.  At low resolution (even with the old 47 cm Swedish
Solar Telescope on La Palma) magnetograms show objects that look like
round flux tubes (Fig.~\ref{stokes_images}, left image).  The new
telescope with twice the diameter will reveal a much more complex
structure (center image), while the simulations show that the actual
situation is even more complex (right image).  The topology of the
observed field may be dramatically altered at low resolution.  Hence,
high resolution observations are crucial for learning the actual
solar magnetic field structure.

Stokes profiles are used to determine the vector magnetic field at
the solar surface.  Figure~\ref{stokes_profiles} shows the stokes
profiles for FeI 6302 as would be observed by the 97 cm Swedish Solar
Telescope on La Palma with perfect seeing (thick lines).  The light
gray lines show the profiles from the 9 individual 25 km$^2$ pixels
covering a 75 $\times$ 75 km region around the central point.  The
observed profile amplitudes are significantly degraded from what
would be seen with infinite resolution.  Observers should take this
as a warning in interpreting their measured profiles.

\subsection{Flux Tubes}

\begin{figure*}[!htb]
\centerline{\psfig{file=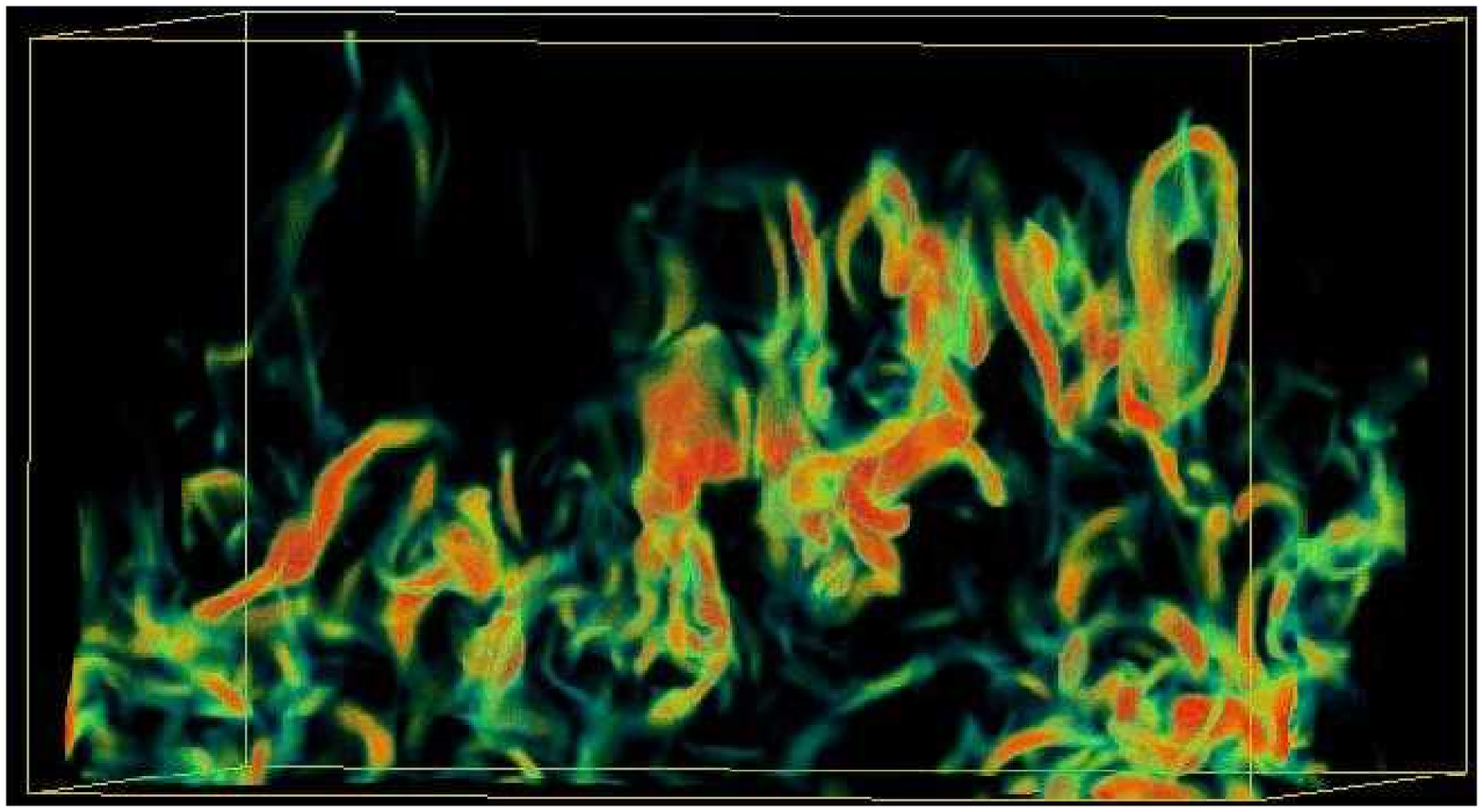,width=13cm}}
\caption{Strong magnetic field concentrations viewed from 
the side.  The visible surface is about 1/7 down from the top.  Are
the small loops with tops above the surface ``flux tubes''?}
\label{fluxtubes_side}
\end{figure*}

\begin{figure*}[!htb]
\centerline{\psfig{file=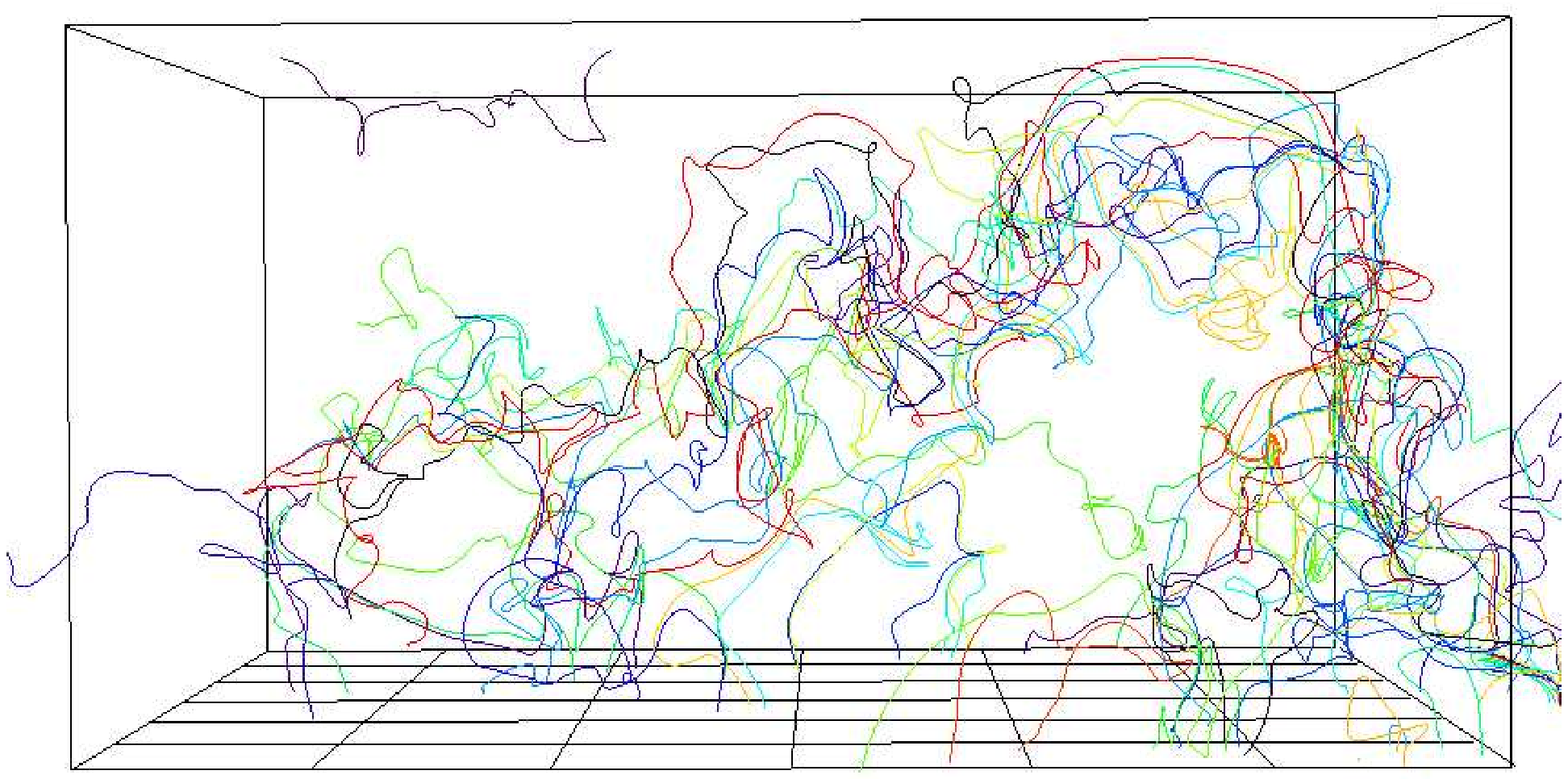,width=\linewidth}}
\caption{The same snapshot as in the previous figure, but now 
showing individual magnetic field lines.  The magnetic field 
lines go into and out of individual strong field concentrations 
and connect over large regions of the simulation domain.
}
\label{fieldlines_side_1}
\end{figure*}

How relevant is the concept of a ``flux tube'' for the weak,
incoherent magnetic fields in the quiet Sun.  Figure
\ref{fluxtubes_side} shows strong magnetic field concentrations from
a snapshot of the 1G simulation.  There appear to be small magnetic
loops extending up through the surface layers (about 1/7 of the
distance down from the top).  However, figure \ref{fieldlines_side_1}
shows that the magnetic field lines go in and out of these loops and
connect several of them.  Thus, what might look like an individual
flux tube is really part of a larger structure.

\begin{figure*}[!htb]
\centerline{
\psfig{file=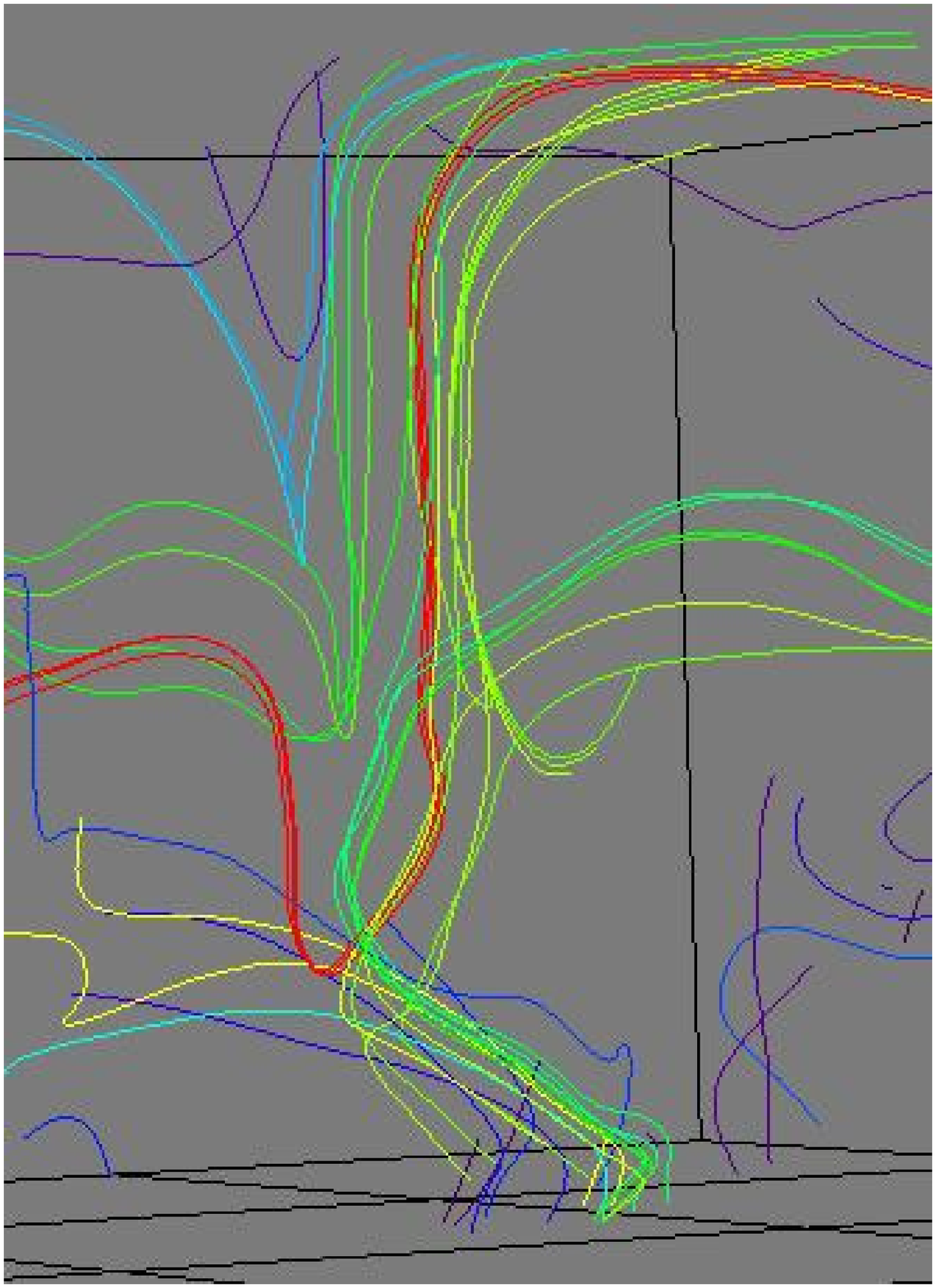,width=4.0 cm}
  \quad
\psfig{file=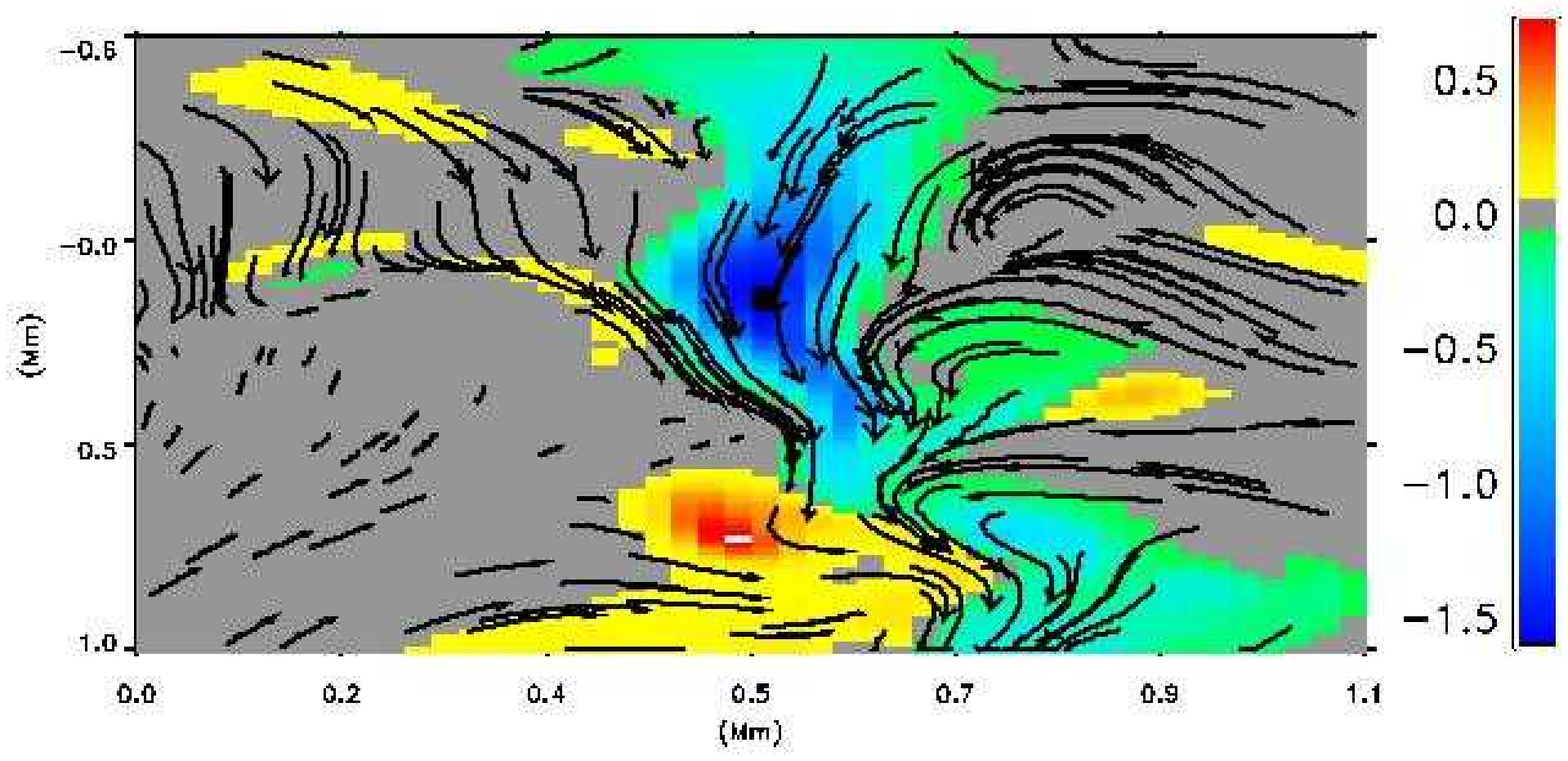,width=9.0 cm}
}
\caption[]{
  \hspace*{-2.7cm}
  \parbox[t]{5.5cm}{
  \vspace*{0.5pc}
(a) Magnetic field lines through a 1.5 kG ``flux tube'' that forms 
and gets partially evacuated in the 30G simulation.  Near the surface 
the magnetic field lines form a flux tube bundle.  Below the 
surface the field lines diverge in several different directions.
}
  \quad
  \parbox[t]{6.5cm}{
  \vspace*{0.5pc}
(b)Image of ``flux tube'' vertical magnetic field with fluid velocity 
vectors in the x-z plane. The tube is in the process of being
evacuated by downflowing fluid.  The density is less than its 
surroundings at and above the surface but greater than its 
surroundings below the surface.
}}
\label{fluxtube_30G}
\end{figure*}

Sometimes, magnetic flux gets concentrated at the vertices of
intergranular lanes and forms a ``flux tube''
(fig.~\ref{fluxtube_30G}).  The ``flux tube'' is cooler than its
surroundings and is being evacuated by downflows, which are strongest
at its periphery, in the intergranule lanes leading into the vertex.
Near the surface and above the density inside the ``flux tube'' is
already less than its surroundings, but at greater depth the density
is still higher than its surroundings.   Notice, that the individual
magnetic field lines that are collected into the ``flux tube''
connect to several different locations below the surface.

\section{Surface Dynamo?}

For a dynamo to work there must be magnetic field amplification by
stretching and twisting, diffusion to reconnect magnetic field lines
and alter their topology, and recirculation to continue the process.
Diverging upflows sweep the fluid into the downflows and concentrate
the magnetic flux, but do not give rise to dynamo action.  Vortical
downdrafts stretch and twist the magnetic field to amplify  it (as
can be seen in the highly twisted magnetic field lines in
fig.~\ref{fieldlines_side_1}).  Resistive diffusion  allows the
magnetic field to diffuse through the fluid and to reconnect.  The
crucial question is whether the field is recirculated to be
continually amplified.

Boussinesq simulations in a closed domain exhibit convection driven
local dynamo action even in the absence of rotation and shearing
motions (Cattaneo 1999, Emonet \& Cattaneo 2001).  The big difference
between our simulation and those of Cattaneo and Emonet is that our
simulation is stratified and has open boundaries.  Magnetic flux is
advected in through the bottom and can also be carried out through
the bottom.  At the top the field tends toward a potential field, so
magnetic flux can also escape that way.  The crucial question is how
much recirculation there is within the near surface layers of the
solar convection zone.

\begin{figure}[!htb]
\centerline{\psfig{file=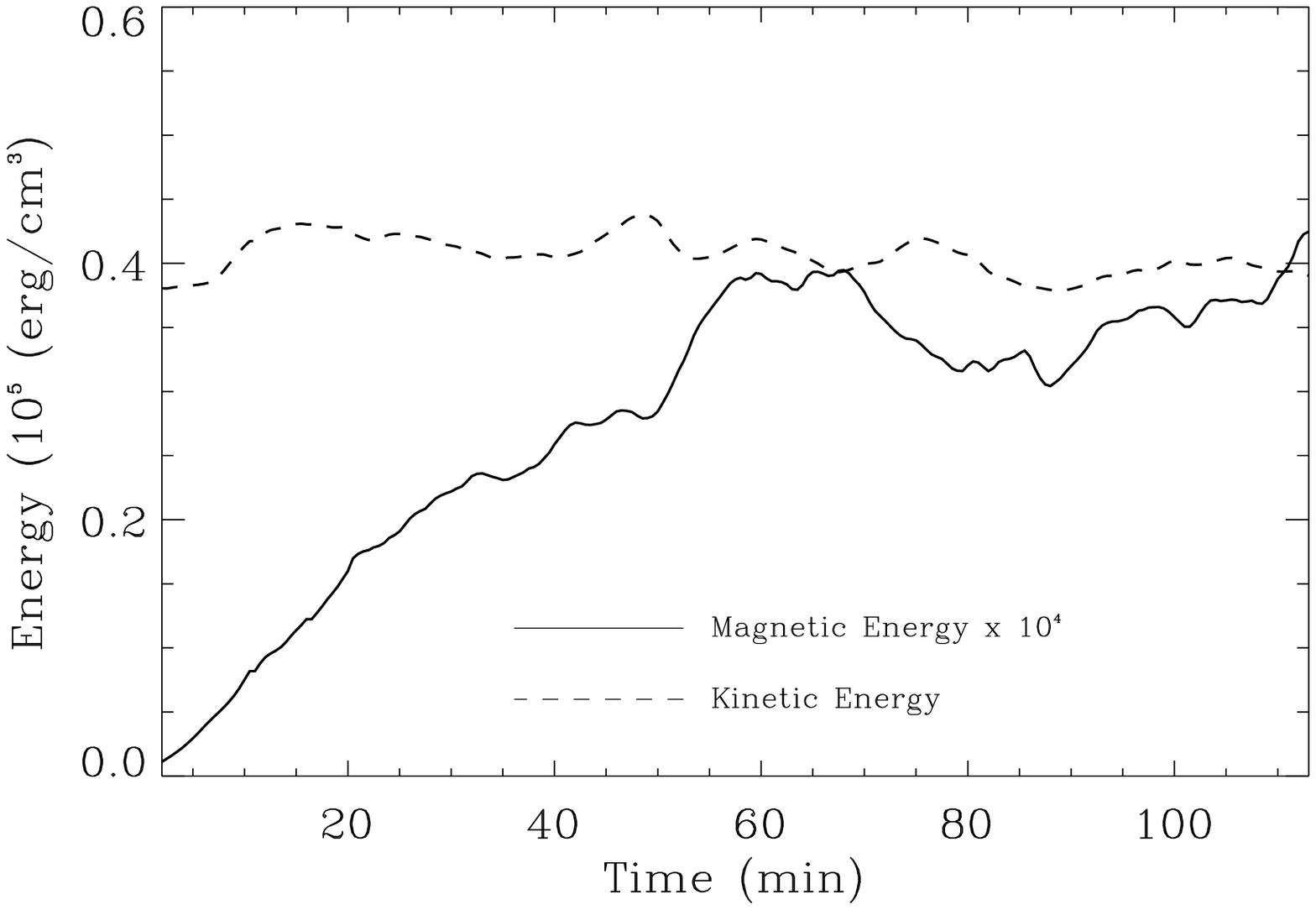,width=8cm}}
\caption[]{
  The magnetic energy increases linearly rather than exponentially, 
  indicating that this is not dynamo action but one pass flux 
  concentration and amplification.  The magnetic energy saturates 
  at a small fraction of the kinetic energy in about a turnover time 
  for the largest (mesogranule) scale in the computational domain.  
  (1G seed field simulation)
}
\label{magnetic_energy}
\end{figure}

\begin{figure*}[!htb]
\vspace*{0.5cm}
\centerline{\psfig{file=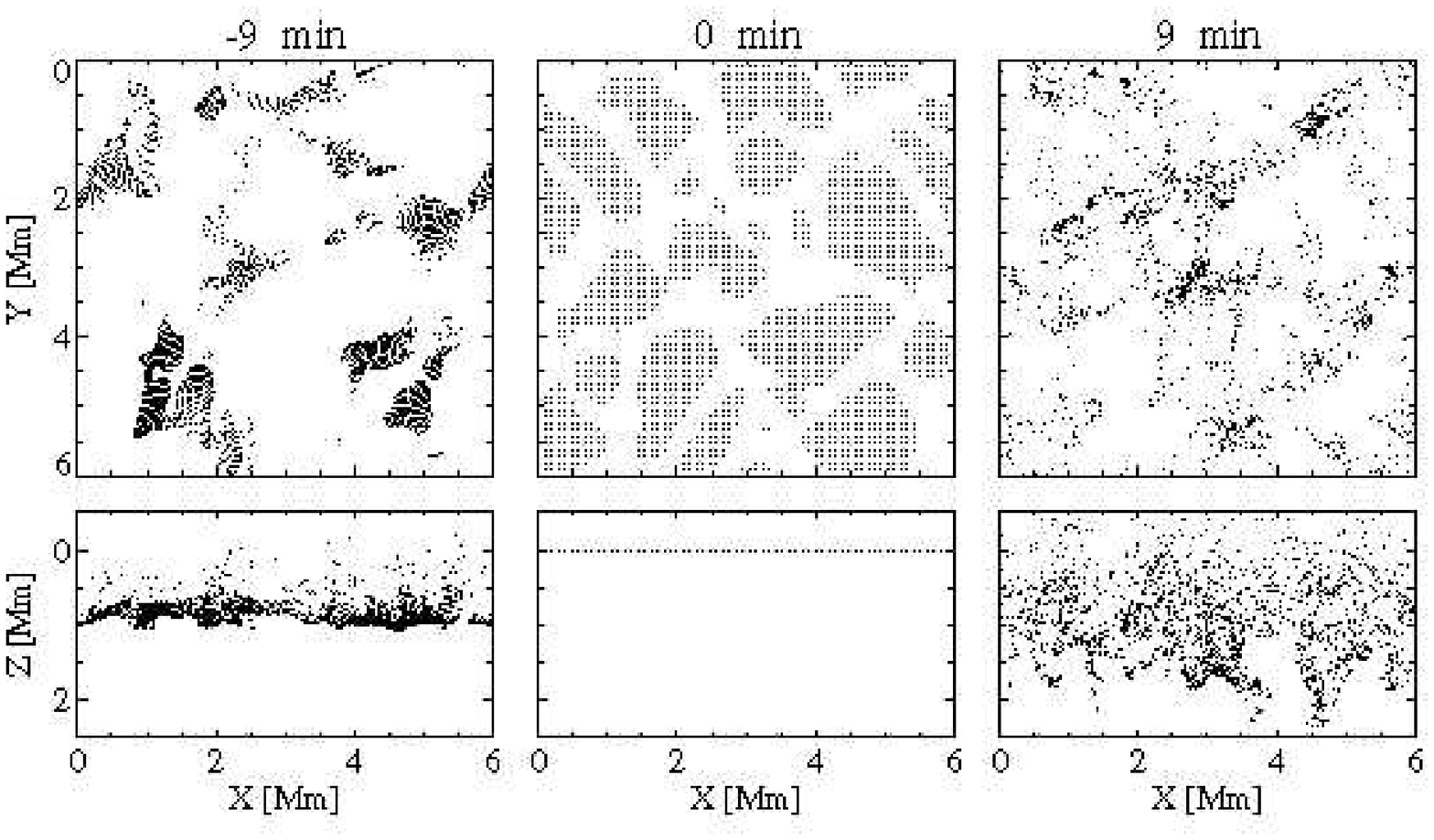,width=\linewidth}}
\vspace*{0.1cm}
\caption{Fluid moving up at the surface and its location 9 min. earlier and 
later.  Little of the fluid rising in the granules at the surface was near
the surface 9 min. earlier.  It comes from a 
small fraction of the volume at depth because the upflows are 
diverging and most ascending fluid turns over within one scale height and 
heads back down in a downdraft in order to conserved mass.  The fluid 
that does reach the surface mostly descends in the fast downdrafts.  
Only a little gets recirculated and was close to the surface 9 min. 
earlier or remains close to the surface after 9 min.
}
\label{parcel_trace}
\end{figure*}

The magnetic energy grows linearly rather than exponentially and
saturates at a small fraction of the kinetic energy
(Fig.~\ref{magnetic_energy}), because
the magnetic field typically passes only once through this region of
concentration and amplification by small scale convective flows.  Unlike 
a Boussinesq fluid, the Sun is highly stratified and convection is
asymmetric, with slow, nearly laminar, diverging upflows and fast,
highly turbulent downflows.  There is only a little local
recirculation near the surface.  Most of the fluid reaching the near
surface layers turns over into downflows and is transported towards
the bottom of the convection zone.  This is illustrated by figure
\ref{parcel_trace} which shows the history of fluid parcels that are
moving upward at the surface at one instant.  Nine minutes earlier
they were mostly at a depth of 1 Mm and occupied a very small
fraction of the horizontal plane.  They occupied only a small
fraction of the area because most ascending fluid must turn over
within a scale height in order to conserve mass, so only a small
fraction of the fluid at depth reaches the surface.  They all come
from approximately the same depth (1 Mm) because they were in upflows
with fairly similar velocities.  Hardly any of them came from near
the surface, an indication of the lack of local recirculation.  Nine
minutes later most of the fluid has left the surface region, many in
narrow downdrafts, and some has even descended to the bottom of the
computational domain, because downflows are fast, converging and
turbulent.  Hardly any are left at or above the surface.  Only a
small portion of the fluid is recirculated to the near surface
layers.  This property is a consequence of the finite physical
separation between the turbulent downdrafts and the small sub-volumes
of the ascending flow that manage to reach the surface, and does not
depend to any significant extent on the Reynolds number.  Due to the
lack of local recirculation, the magnetic energy grows linearly
rather than exponentially because the magnetic field typically passes
only once through the surface region where it is concentrated and
amplified by small scale convective flows.

\begin{figure}[!tb]
\centerline{ \psfig{file=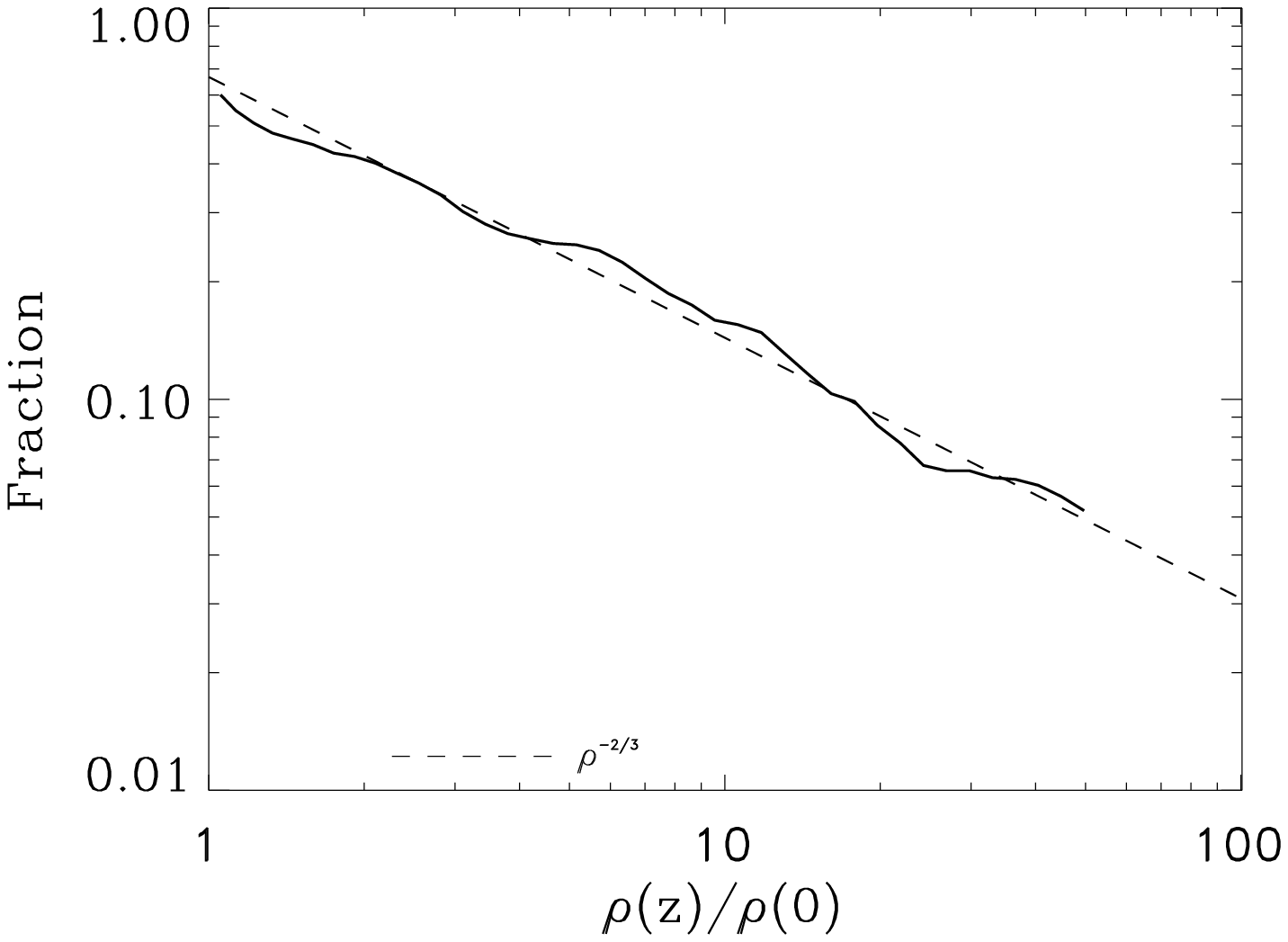,width=8cm} }
\caption[]{
Fraction of fluid reaching the surface from any depth scales as 
$(\rho/\rho_{\mbox{surface}})^{-2/3}$.  
}
\label{pumping1}
\end{figure}

\begin{figure}[th]
\centerline{ \psfig{file=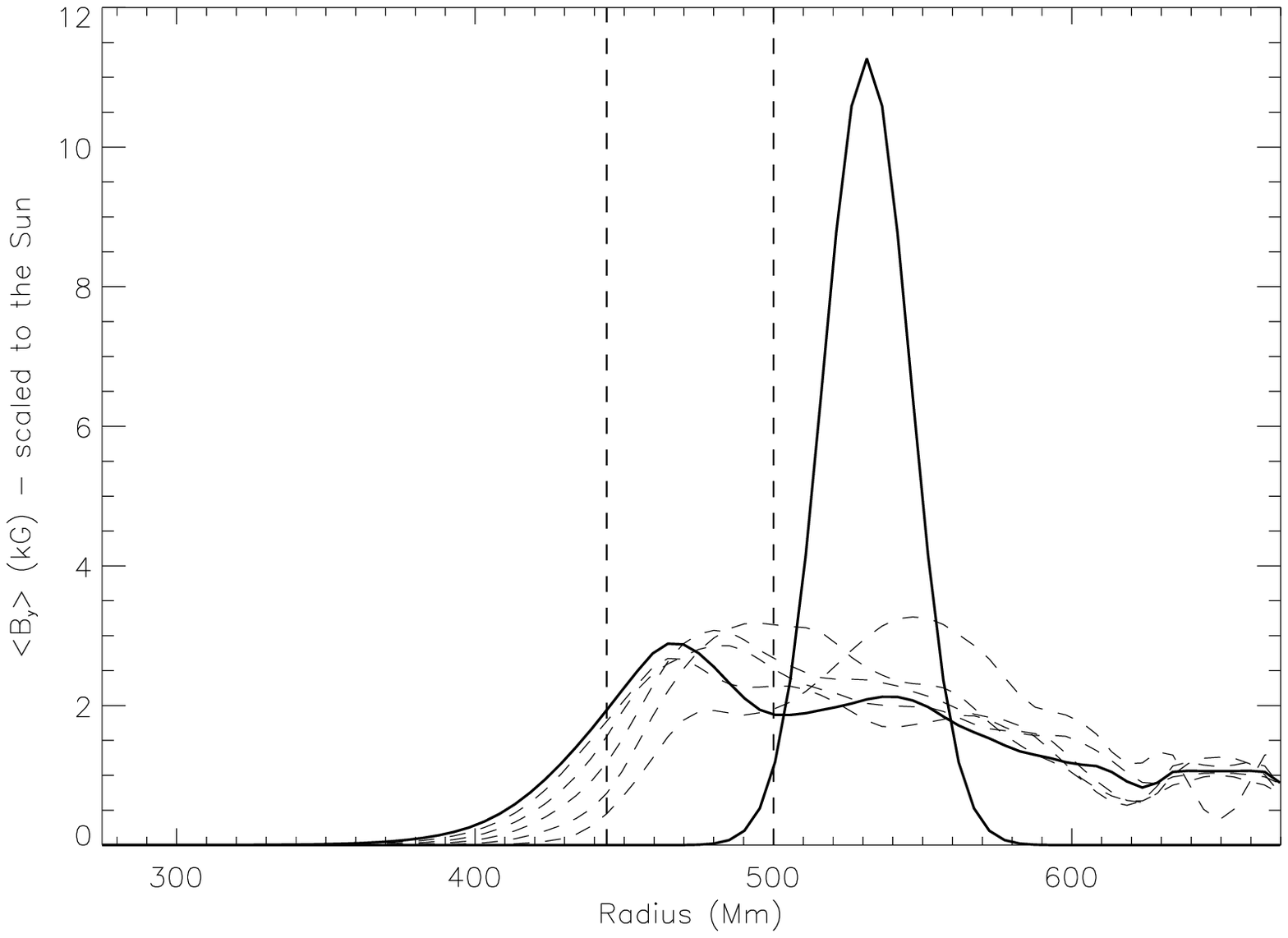,width=8cm} }
\caption[]{
Magnetic flux initially in a thin layer in the convection zone 
gets redistributed throughout the convection zone, with most in the deeper 
layers (Nordlund et al. 2000).  The right portion is the convection 
zone, the two vertical lines mark the undershoot layer, and the 
left portion is stably stratified.}
\label{pumping2}
\end{figure}

Mass conservation requires that at any depth most upflowing fluid
must turn over and head back down within about a scale height.
Hence, only a tiny fraction of fluid starting up from the bottom of
the convection zone makes it all the way to the surface.  By tracing
individual fluid parcels in time, we have found that the fraction
reaching the surface from a depth where the density is $\rho(z)$
decreases as $(\rho(z)/\rho(\mbox{surface}))^{-2/3}$
(fig.~\ref{pumping1}).  The weak incoherent magnetic field, unlike
the strong coherent active region flux, is dominated by drag and
advected by the fluid.  Hence, most of the magnetic field will be
located in the deeper parts of the convection zone.  This has been
verified by numerical simulations
(Dorch \& Nordlund 2001, Tobias et al. 2001).  Magnetic field placed
anywhere in the convection zone gets distributed throughout the
convection zone and into the overshoot layer (fig.~\ref{pumping2}).
The field strength increases with depth and its maximum lies in the
overshoot layer, but most of the magnetic flux and energy is inside
the convection zone, concentrated near the bottom.  (In the Sun the
tachocline is much thinner than it is in this toy simulation, so the
amount of magnetic energy and flux in it is a small fraction of the
total magnetic energy and flux.)  The magnetic energy and flux near
the surface is likewise small compared to the total magnetic energy
and flux because it gets carried down to deeper layers by the
downflows and only a little is brought back to the surface by the
upflows.

Is there a surface dynamo?  In the Sun, unlike simulations with
closed boundaries and without stratification, there is little local
surface recirculation.  The recirculation is global.  Surface
magnetic fields are carried toward the bottom of the convection zone
by downdrafts.  The time scale at the bottom of the convection zone
is long, months.  Only a tiny fraction of fluid ascending from the
bottom of the convection zone reaches the surface.  Hence, much more
flux and magnetic energy resides in the deep convection zone than
near the surface and the energy added to flux that visits the surface
is tiny compared to the global magnetic energy.  These results lead
us to conclude that there is no localized surface dynamo.  Rather,
there is a global dynamo in which a small fraction of weak,
incoherent fields residing inside the convection zone is dragged to
the surface, where it is shredded, concentrated, stretched and
twisted by the small scale surface convective motions into the
observed, incoherent, intermittent continually emerging, small scale
surface field.  This one-pass surface dynamo action on granular and
mesogranular (and probably supergranular) scales would distinguish
these incoherent fields from the active region magnetic fields which
represent another component of the global dynamo: strong, coherent
fields that are buoyant and sufficiently large scale to feel the
coriolis force, differential rotation and meridional circulation.

\acknowledgments

This work was supported in part by NASA grant NAG 5 9563 and NSF
grant AST 9819799, and by the Danish Research Foundation through its
establishment of the Theoretical Astrophysics Center.  The
calculations were performed at National Center for Supercomputing
Applications (which is supported by NSF), Michigan State University,
and DCSC, Denmark.  Their support is greatly appreciated.

%
%


\begin{references}

\reference Cattaneo, F. 1999, \apjlett, 515, L39

\reference Dorch, S.B.F. \& Nordlund, \AA. 2001, \aap, 365, 562

\reference Emonet, T. \& Cattaneo, F. 2001, \apjlett, 560, L197

\reference Hagenaar, H.J. 2001, \apj, 555, 448

\reference Harvey, K.L. \& White, O.R. 1999, \apj, 515, 812

\reference Nordlund, \AA., Dorch, S.B.F., \& Stein, R.F. 2000, J. of
  Astrophys. \& Astron., 21, 307

\reference Stein R.F. \& Nordlund {\AA}. 2000, Solar Phys., 192, 91

\reference Tobias, S.M., Brummell, N.H., Clune, T.L. \& Toomre, J., 2001,
  \apj, 549, 1183

\end{references}
%

%

\end{document}